%% file: main.tex
\documentclass[conference,letterpaper]{IEEEtran}
\IEEEoverridecommandlockouts
\usepackage[T1]{fontenc}
\usepackage[utf8]{inputenc}
\usepackage{times}
\usepackage{amsmath,amssymb}
\usepackage{graphicx}
\usepackage{booktabs}
\usepackage{cite}
\usepackage{stfloats}
\graphicspath{{figures/}}

\title{A Two-Stage Training Protocol for Indoor Wall\\
Geometry Inference from Sparse 5G RSSI\\
Observations}
\input{authors}

\begin{document}
\maketitle
\input{sections/00_abstract}
\input{sections/01_introduction}
\input{sections/02_related_work}
\input{sections/03_problem_dataset}
\input{sections/04_architecture}
\input{sections/05_training}
\input{sections/06_results}
\input{sections/07_discussion}
\input{sections/08_conclusion}
\input{references/references}

\end{document}

%% file: authors.tex
\author{%
\IEEEauthorblockN{Yongyuan Hu\thanks{Corresponding author: Yongyuan Hu.}}
\IEEEauthorblockA{Swanson School of Engineering\\
University of Pittsburgh\\
Pittsburgh, PA, USA\\
derrick.hu@pitt.edu}
\and
\IEEEauthorblockN{Huimin Wang}
\IEEEauthorblockA{College of Computer Science\\
Sichuan University\\
Chengdu, China\\
huimin.xxtz@gmail.com}
}

%% file: sections/00_abstract.tex
\begin{abstract}
Inferring indoor partition geometry from sparse opportunistic radio measurements enables environment mapping without dedicated hardware. We identify a systematic encoder collapse in Transformer-based architectures under sparse spatial scatter. We therefore propose a two-stage training protocol that resolves it: stage 1 trains an encoder to reconstruct the full received signal strength indicator (RSSI) heatmap as a pretext task; stage 2 optimises an independent wall decoder using the frozen encoder representations, followed by joint end-to-end fine-tuning with a regularised composite loss. Evaluated on 20,000 Sionna RT ray-traced samples spanning seven topologies at 3.5~GHz, the convolutional neural network (CNN) U-Net achieves Wall $F_1$ of 0.879 in single-stage training, and 0.854 under the two-stage protocol; a Transformer encoder reaches $F_1$ of 0.792. Omitting stage 1 causes the Transformer encoder to collapse to a constant output ($F_1 = 0.166$ regardless of observation count), whereas CNNs tolerate single-stage training. We introduce the first RSSI-to-wall benchmark dataset.
\end{abstract}

%% file: sections/01_introduction.tex
\section{Introduction}
Accurate indoor partition geometry underpins 5G base-station placement, beam management, digital twin construction, and radio environment map generation~\cite{ref1}. However, conventional mapping pipelines---LiDAR simultaneous localization and mapping (SLAM), RGB-D reconstruction, terrestrial laser scanning---impose significant hardware cost and operational overhead~\cite{ref2,ref3}, and for a large fraction of older buildings no usable floor-plan data exists at all. Camera-based approaches trade a geometry gap for a privacy constraint~\cite{ref4}.

5G downlink signals already present in the environment provide an alternative acquisition path. Received signal strength indicator (RSSI) is a scalar aggregate of the full channel impulse response; wall interactions encode as shadow perturbations on top of free-space path loss. Inverting sparse RSSI measurements to recover wall geometry is a highly nonlinear mapping problem. Although deep learning is a natural tool for this task, direct end-to-end training on sparse spatial inputs is prone to failure due to insufficient gradient diversity during early training.

We address this with a two-stage curriculum: the first stage learns a dense RSSI heatmap reconstruction as a spatial pretext objective, forcing the encoder to build propagation-aware features before the sparse wall supervision is introduced; the second stage trains a separate wall decoder on the frozen encoder output, after which all parameters are jointly refined with a composite loss. We systematically evaluate this decomposition against single-stage training and discover that the pretext stage is essential for Transformer encoders, which collapse to degenerate constant outputs without it---a failure mode that CNN architectures do not exhibit.

Contributions:
\begin{itemize}
\item The first RSSI-to-wall benchmark dataset, comprising 20,000 ray-traced indoor scenes across seven wall configurations at 3.5~GHz.
\item A systematic evaluation of seven architectures under both single-stage and two-stage training across five observation densities, identifying which regime favours which model.
\item A mechanistic finding: Transformer encoders irreversibly collapse to constant outputs under single-stage training with sparse spatial inputs ($F_1 = 0.166$ regardless of the number of observations); the two-stage protocol resolves this, while CNN architectures bypass it entirely.
\end{itemize}

%% file: sections/02_related_work.tex
\section{Related Work}
Existing geometry acquisition methods each carry practical drawbacks: LiDAR and RGB-D pipelines demand dedicated sensors and non-trivial deployment effort~\cite{ref2,ref3}, while scan-to-building-information-modelling (BIM) conversion remains unavailable for the majority of older building stock~\cite{ref5}. Integrated sensing and communication (ISAC) offers a lighter-weight alternative by extracting geometric cues from wireless signals~\cite{ref1}, yet angle-of-arrival techniques such as SpotFi~\cite{ref6} rely on multiple-input multiple-output (MIMO) arrays without recovering full room geometry, and RSS fingerprinting methods require dense labelled site surveys.

Recent work has extended radio-based sensing to received signal strength (RSS)-only geometry inference. Lederboer et al.~\cite{ref7} detect indoor walls from RSS measurements using machine learning; Kim et al.~\cite{ref8} recover wall boundaries from WiFi RSSI using an inverse visibility algorithm validated in real indoor environments, while Chen and Chen~\cite{ref9} jointly reconstruct a 3D propagation environment and radio map from received signal strength via diffraction-aware deep learning. Gaussian process regression~\cite{ref10} and neural processes~\cite{ref11,ref12} address sparse spatial inference but not wall-geometry inversion. RadioUNet~\cite{ref13} and deep completion autoencoders~\cite{ref14} learn radio maps conditioned on a known building layout. Our work solves the inverse problem of inferring the layout from the radio measurements.

Masked autoencoders~\cite{ref15} motivate our use of an intermediate dense reconstruction task to regularise representation learning. Self-supervised pretraining with dense reconstruction objectives has demonstrated strong transfer in the vision domain~\cite{ref15} and beyond. Our two-stage protocol applies this idea to sparse spatial regression, with dense heatmap recovery serving as the pretext objective; self-supervised objectives have similarly shown promise in RF sensing domains~\cite{ref16,ref17}. Two-stage and multi-task training protocols have been applied in radar sensing~\cite{ref1} and radio map estimation~\cite{ref13}, but without the explicit frozen-encoder stage that we identify as critical for attention-based encoders under sparse observations.

%% file: sections/03_problem_dataset.tex
\section{Problem Formulation \& Dataset}
Consider a 2D indoor room of size $10\times10$~m. A transmitter at known position $\mathbf{p}_{\mathrm{tx}}\in[0,1]^2$ (normalised) emits 5G New Radio (NR) signals at 3.5~GHz. A mobile receiver collects $N$ sparse observations $\mathcal{X}=\{(\mathbf{x}_i,r_i)\}_{i=1}^{N}$, where $\mathbf{x}_i\in[0,1]^2$ is position and $r_i\in[0,1]$ is normalised RSSI. The task is to infer wall occupancy $\hat{W}\in[0,1]^{64\times64}$, where ground-truth $W_{ij}=1$ indicates a wall at grid cell $(i,j)$. Ground-truth occupancy maps are obtained by rasterising wall segment centrelines onto the $64\times64$ grid with a Gaussian kernel ($\sigma=0.156$~m, one grid cell width), yielding soft probabilistic targets that provide smooth gradients for MSE-based optimisation.

Sionna RT~\cite{ref18} and similar ray-tracing simulators have been validated for sub-6~GHz indoor propagation~\cite{ref19}, providing a physically accurate simulation basis for our dataset. We generated 20,000 ray-traced indoor scenes using Sionna RT at 3.5~GHz across seven wall topology types (Single Short, Single Long, Cross, Cross Extended, Parallel, Triple, and T-shape), incorporating ITU-R standard material properties~\cite{ref20}, following channel models specified in 3GPP TR 38.901~\cite{ref21}. Within each scene, a transmitter is placed at a random position, and 600 RSSI samples are collected at uniformly random receiver positions. The dataset is split into 16,000 training, 2,000 validation, and 2,000 test scenes with balanced topology distribution. A separate path-based test set of 200 scenes uses a random-walk trajectory with $N=48$ observations, simulating a smartphone user traversing the room.

We evaluate the models using three complementary metrics: (i) mean squared error (MSE) between predicted and ground-truth wall occupancy maps; (ii) the intersection over union (IoU) at threshold 0.5; and (iii) Wall $F_1$ at threshold 0.5. $F_1$ is our primary metric as it balances precision (avoiding false walls) and recall (detecting true walls) in the presence of class imbalance (walls occupy 3\%--10\% of cells).

%% file: sections/04_architecture.tex
\section{Architecture}
\subsection{Spatial Scatter Encoding}
Each observation $(\mathbf{x}_i,r_i)$ is augmented with the transmitter (TX) position to form a 5-dimensional token $[\mathbf{x}_i;\,r_i;\,\mathbf{p}_{\mathrm{tx}}]$. A learned scatter layer soft-assigns observation tokens to a $4\times4$ spatial grid via attention-weighted aggregation, producing a fixed $16\times64$ grid feature tensor that serves as the encoder input.

\subsection{Transformer Encoder}
The encoder is a 6-layer multi-head Transformer ($d_{\mathrm{model}}=512$, $h=8$, dropout $=0.15$) with rotary position embeddings on the $4\times4$ grid tokens. Its output is reshaped into a $64\times4\times4$ spatial latent code $\mathbf{z}$. For the CNN U-Net baseline, the same $16\times64$ feature matrix is processed by a convolutional U-Net encoder under the identical two-stage protocol.

\subsection{Heatmap Decoder (Stage 1)}
A lightweight transposed-convolution decoder progressively upsamples $\mathbf{z}$ from $64\times4\times4$ to $1\times64\times64$ via four bilinear interpolation stages, yielding the reconstructed RSSI heatmap $\hat{H}$. Stage 1 (the heatmap pretext) minimises the mean squared reconstruction error: $\mathcal{L}_1=\|H-\hat{H}\|^2$.

\subsection{Wall Decoder (Stage 2) and End-to-End Fine-Tuning}
An independent transposed-convolution decoder of identical structure maps $\mathbf{z}$ to soft wall occupancy $\hat{W}$. Stage 2 freezes the encoder and trains only the wall decoder: $\mathcal{L}_2=\operatorname{MSE}(\hat{W},W_{\mathrm{GT}})$. After stage 2 converges, a joint fine-tuning phase adapts all parameters end-to-end with $\mathcal{L}_{\mathrm{total}}=0.5\,\mathcal{L}_1+\mathcal{L}_2$. The composite loss weight $\lambda=0.5$ was selected by grid search over $\{0.1,0.3,0.5,0.7,0.9\}$ on the validation set. Performance was stable across $\lambda\in[0.3,0.7]$, indicating that the heatmap reconstruction term functions as a regulariser rather than a competing objective during joint fine-tuning. The full Transformer 2S model has approximately 12M parameters; CNN U-Net 2S has 6.5M; multi-layer perceptron (MLP) 2S has 1.2M. All 2S results reported in this paper include the joint fine-tuning phase.

%% file: sections/05_training.tex
\section{Baselines and Training Details}
We evaluate seven models: Transformer 2S (proposed two-stage Transformer, primary model); CNN U-Net 2S (U-Net encoder, same two-stage protocol); CNN U-Net 1S (U-Net, single-stage, no heatmap pretext); MLP 2S (two-stage MLP, no spatial structure); Trans. No-Spatial (Transformer without $4\times4$ spatial scatter, global mean pooling instead); Transformer 1S (single-stage ablation, stage 1 removed); and GP + Gradient (Gaussian process RSSI interpolation followed by image-gradient wall detection~\cite{ref10}).

\begin{figure*}[!b]
\centering
\includegraphics[width=\textwidth]{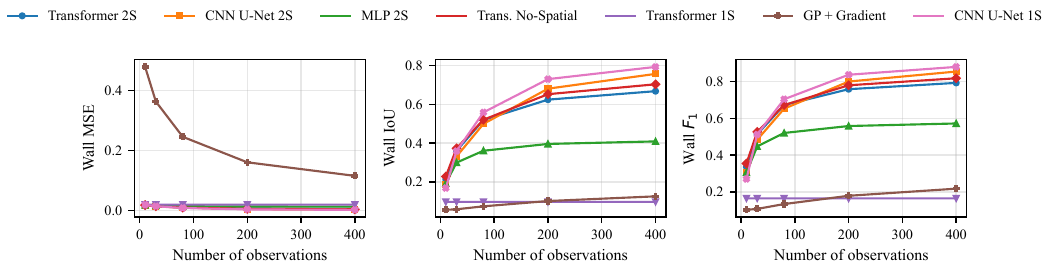}
\caption{Wall reconstruction metrics vs.\ number of observations (validation set). Transformer 1S is flat at $F_1=0.166$ regardless of $N$ (encoder collapsed).}
\label{fig:metrics}
\end{figure*}

\input{tables/table1_main}

All neural models are optimised with AdamW ($\beta_1=0.9$, $\beta_2=0.999$, $\varepsilon=10^{-8}$, weight decay $\lambda_{\mathrm{wd}}=10^{-4}$) under a cosine annealing learning-rate schedule with warm restarts ($T_0=20$ epochs). Stage 1 runs for 100 epochs; stage 2 for 80 epochs with the encoder frozen; joint fine-tuning then adapts all parameters for 40 epochs with learning rate reduced by a factor of 10. Training uses a batch size of 64 on a single NVIDIA A100 GPU. At inference, observation positions are randomly sub-sampled from the 600 available positions to simulate sparse collection scenarios. For path-based evaluation, observations follow a random-walk trajectory constrained to the room boundary with step length 0.3~m and a $180^\circ$ field of view, yielding $N=48$ spatially correlated samples.

%% file: tables/table1_main.tex
\begin{table*}[!t]
\caption{Wall Reconstruction Performance (Validation Set). SSIM omitted from table and figures (near-zero for all neural models).}
\label{tab:main}
\centering
\setlength{\tabcolsep}{3.2pt}
\footnotesize
\begin{tabular}{l*{15}{r}r}
\toprule
Method & \multicolumn{3}{c}{$N=10$} & \multicolumn{3}{c}{$N=30$} & \multicolumn{3}{c}{$N=80$} & \multicolumn{3}{c}{$N=200$} & \multicolumn{3}{c}{$N=400$} & Latency (ms)\\
& MSE & IoU & $F_1$ & MSE & IoU & $F_1$ & MSE & IoU & $F_1$ & MSE & IoU & $F_1$ & MSE & IoU & $F_1$ &\\
\midrule
Transformer 2S & .0193 & .213 & .337 & .0157 & .361 & .513 & .0108 & .520 & .672 & .0076 & .625 & .758 & .0065 & .669 & .792 & 16\\
CNN U-Net 2S & .0194 & .188 & .304 & .0156 & .331 & .484 & .0109 & .500 & .652 & .0058 & .682 & .799 & .0039 & .758 & .854 & 18\\
MLP 2S & .0194 & .192 & .306 & .0171 & .300 & .448 & .0152 & .361 & .520 & .0142 & .396 & .558 & .0138 & .409 & .572 & 3\\
Trans. No-Spatial & .0188 & .228 & .355 & .0150 & .373 & .526 & .0104 & .521 & .672 & .0065 & .653 & .779 & .0051 & .704 & .817 & 14\\
Transformer 1S & .0205 & .097 & .166 & .0205 & .097 & .166 & .0205 & .097 & .166 & .0205 & .097 & .166 & .0205 & .097 & .166 & 9\\
GP + Gradient & .4782 & .055 & .103 & .3629 & .058 & .109 & .2458 & .074 & .135 & .1611 & .101 & .180 & .1161 & .125 & .218 & --\\
CNN U-Net 1S & .0197 & .168 & .271 & .0150 & .359 & .512 & .0092 & .559 & .703 & .0046 & .731 & .836 & .0031 & .794 & .879 & 9\\
\bottomrule
\end{tabular}
\par\smallskip
\footnotesize -- indicates latency not measured for non-neural baseline (GP + Gradient).
\end{table*}

%% file: sections/06_results.tex
\section{Experiments and Results}
\subsection{Main Results}
Table~\ref{tab:main} reports Wall MSE, IoU, and $F_1$ on the 2,000-scene validation set at $N\in\{10,30,80,200,400\}$. The most striking result is the complete collapse of Transformer 1S: $F_1=0.166$ at every observation count, equivalent to a near-constant baseline output. This collapse is irreversible---increasing $N$ from 10 to 400 produces no improvement, indicating that without the heatmap pretext the Transformer encoder fails to learn any spatially useful representation. In contrast, CNN U-Net 1S reaches $F_1=0.879$ at $N=400$, showing that CNN architectures can learn wall geometry in a single stage. Among two-stage models, CNN U-Net 2S attains $F_1=0.854$ at $N=400$, with Transformer 2S at 0.792. Trans. No-Spatial slightly outperforms Transformer 2S at $N=400$ (0.817 vs.\ 0.792), a counter-intuitive result analysed further in Section~\ref{sec:discussion}. MLP 2S saturates at $F_1=0.572$, confirming the value of spatial structure. GP + Gradient improves with $N$ but remains below 0.22 at $N=400$, reflecting fundamental limitations of gradient-based edge detection on smooth interpolated RSSI surfaces.

Per-sample inference latency, measured on the same GPU with batch size 1, ranges from 3~ms (MLP 2S) to 16~ms (Transformer 2S) and 18~ms (CNN U-Net 2S), satisfying the sub-100~ms latency budget typical of near-real-time radio environment mapping pipelines.

\subsection{$F_1$ vs. Observation Count}
Fig.~\ref{fig:metrics} plots Wall $F_1$ as a function of $N$ for all models. Two-stage Transformer and CNN U-Net both improve steadily with $N$, while Transformer 1S is flat at 0.166 across all regimes, confirming irreversible encoder collapse. The $F_1$ gain from $N=10$ to $N=400$ is 0.455 for Transformer 2S and 0.608 for CNN U-Net 1S, highlighting that models with spatial encoding benefit substantially from denser observations. CNN U-Net 1S slightly outperforms its two-stage counterpart at $N\geq200$, consistent with the regime summary in Table~\ref{tab:regimes}. The gap between Transformer 2S and Trans. No-Spatial narrows as $N$ decreases, suggesting the scatter grid is most valuable under extreme sparsity. Training converges stably across all stages; stage 2 reaches 90\% of its final $F_1$ within 30 epochs due to the well-initialised encoder features.

\begin{figure}[!t]
\centering
\includegraphics[width=\columnwidth]{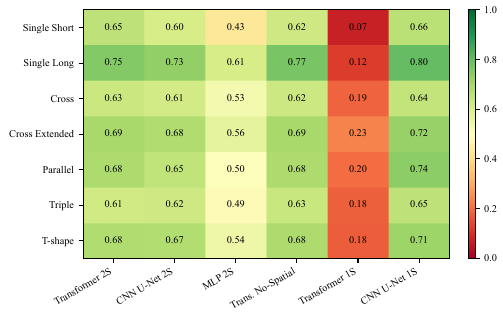}
\caption{Per-topology Wall $F_1$ at $N=80$ (validation set).}
\label{fig:topology}
\end{figure}
\input{tables/table2_regimes}

\subsection{Per-Topology Analysis}
\label{sec:results-topology}
Fig.~\ref{fig:topology} shows $F_1$ broken down by wall topology at $N=80$. Single Long yields the highest $F_1$ across all models; Single Short also tends to perform well, though less consistently. Both benefit from fewer ambiguous intersection points and larger unobstructed RSSI gradients compared to complex layouts. Complex geometries (Triple, Cross Extended) are harder because multiple wall intersections create conflicting shadow gradients that challenge all encoders. T-shape is an intermediate case: the asymmetry of the topology provides a distinctive RSSI footprint that some models---particularly CNN U-Net architectures---exploit effectively. Transformer 1S remains collapsed across every topology, further confirming that the failure mode is not topology-specific but stems from the absence of the heatmap pretext stage.

\subsection{Observation-Regime Summary}
Table~\ref{tab:regimes} aggregates $F_1$ by observation regime. In the low regime ($N\in\{10,30\}$), all models except Transformer 1S cluster near $F_1\approx0.4$, reflecting the fundamental ambiguity of very sparse coverage. In the mid regime ($N\in\{80,200\}$), Transformer 2S and CNN U-Net 2S reach $F_1=0.715$ and 0.726, respectively, while MLP 2S lags at 0.539, confirming the importance of spatial encoding at moderate densities. In the high regime ($N=400$), CNN U-Net variants pull ahead while MLP 2S saturates at 0.572. Trans. No-Spatial (0.817) edges out Transformer 2S (0.792) at high $N$, consistent with the per-count analysis in Table~\ref{tab:main}.

\input{tables/table3_path}

\begin{figure*}[!t]
\centering
\includegraphics[width=\textwidth]{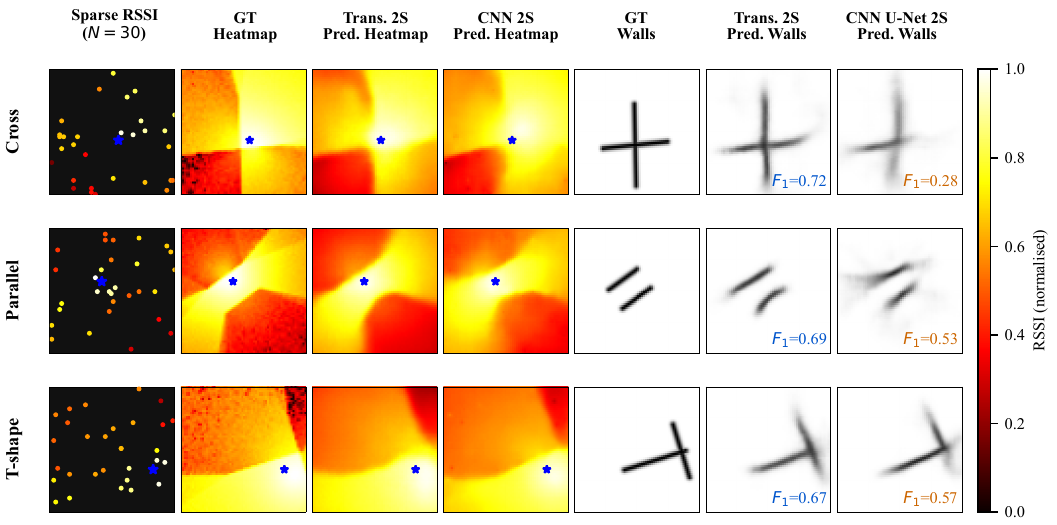}
\caption{Inference pipeline at $N=30$ for three topology types (rows). Columns: sparse RSSI observations (colour = normalised RSSI, blue star = TX); ground-truth RSSI heatmap; Transformer 2S stage-1 predicted heatmap; CNN U-Net 2S stage-1 predicted heatmap; ground-truth wall occupancy map; Transformer 2S stage-2 predicted walls ($F_1$ in blue); CNN U-Net 2S stage-2 predicted walls ($F_1$ in orange). At $N=30$, Transformer 2S outperforms CNN U-Net 2S on all three topologies, demonstrating the advantage of the spatial scatter grid under sparse observations.}
\label{fig:inference}
\end{figure*}

\subsection{Qualitative Results}
Fig.~\ref{fig:inference} illustrates the full inference pipeline at $N=30$ for three topology types (Cross, Parallel, and T-shape), comparing Transformer 2S and CNN U-Net 2S under the same two-stage protocol. Columns 1--4 show the sparse RSSI observations, the ground-truth RSSI heatmap, and the stage-1 predicted heatmaps from each model, confirming that both encoders reconstruct spatially coherent propagation fields from 30 scattered measurements. Columns 5--7 show ground-truth and predicted wall maps: at $N=30$, Transformer 2S consistently outperforms CNN U-Net 2S ($F_1$ gains of 0.44, 0.16, and 0.10 on the three topologies), demonstrating that the spatial scatter grid provides a stronger structural prior under extreme observation sparsity.

\subsection{Path-Based Per-Topology $F_1$}
Table~\ref{tab:path} reports Wall $F_1$ on the path-based test set ($N=48$, random-walk trajectory) broken down by topology. All models are substantially weaker than on the random-sampling validation set; the best model (Trans. No-Spatial) averages $F_1\approx0.24$ across topologies, with a per-topology peak of 0.384 on Cross Extended. Note that Trans. No-Spatial omits the spatial scatter module entirely; its marginal advantage at high $N$ is analysed in Section~\ref{sec:discussion} and reflects a resolution bottleneck in the fixed $4\times4$ grid rather than a limitation of the two-stage protocol. Unlike the random-sampling regime where complex topologies are harder (see Section~\ref{sec:results-topology}), under path-based sampling longer walls intersect more trajectory positions, making them easier to detect: Cross Extended and Parallel outperform Single Short for this reason. Single Long and Triple topologies are excluded because the random-walk trajectory provides insufficient wall coverage for reliable $F_1$ estimation in those layouts.

All neural models degrade gracefully under RSSI noise $\sigma\leq0.20$; the GP baseline collapses at $\sigma>0.05$. Ceiling-mounted TX yields the best $F_1$ across all models; wall-mounted is the worst due to anisotropic shadow zones.

%% file: tables/table2_regimes.tex
\begin{table}[!t]
\caption{Wall $F_1$ by Observation Regime (Validation Set).}
\label{tab:regimes}
\centering
\footnotesize
\setlength{\tabcolsep}{5pt}
\begin{tabular}{lccc}
\toprule
Method & Low (10--30) & Mid (80--200) & High (400)\\
\midrule
Transformer 2S & 0.425 & 0.715 & 0.792\\
CNN U-Net 2S & 0.394 & 0.726 & 0.854\\
MLP 2S & 0.377 & 0.539 & 0.572\\
Trans. No-Spatial & 0.441 & 0.726 & 0.817\\
Transformer 1S & 0.166 & 0.166 & 0.166\\
GP + Gradient & 0.106 & 0.157 & 0.218\\
CNN U-Net 1S & 0.392 & 0.770 & 0.879\\
\bottomrule
\end{tabular}
\end{table}

%% file: tables/table3_path.tex
\begin{table}[!t]
\caption{Per-Topology Wall $F_1$ on the Path-Based Test Set ($N=48$). Single Long and Triple excluded due to insufficient path-trajectory coverage in those topologies. GP + Gradient excluded as GP interpolation requires spatially dense observations and produces unreliable wall gradients under correlated single-trajectory sampling.}
\label{tab:path}
\centering
\scriptsize
\setlength{\tabcolsep}{3pt}
\begin{tabular}{lrrrrrr}
\toprule
Topology & T2S & CNN2S & MLP2S & T-NS & T1S & CNN1S\\
\midrule
Single Short & 0.107 & 0.156 & 0.107 & 0.138 & 0.120 & 0.133\\
Cross & 0.178 & 0.163 & 0.144 & 0.194 & 0.158 & 0.121\\
Cross Ext. & 0.364 & 0.301 & 0.315 & 0.384 & 0.269 & 0.255\\
Parallel & 0.275 & 0.231 & 0.224 & 0.280 & 0.260 & 0.202\\
T-shape & 0.223 & 0.249 & 0.164 & 0.188 & 0.189 & 0.156\\
\bottomrule
\end{tabular}
\end{table}

%% file: sections/07_discussion.tex
\section{Discussion}
\label{sec:discussion}
\textbf{Why two-stage training works.} Wall occupancy maps are sparse binary images whose statistics are largely independent of transmitter position, while RSSI heatmaps are smooth scalar fields carrying explicit spatial information about propagation obstacles. Pretraining the encoder on heatmap reconstruction biases it toward spatially coherent feature representations before the wall decoding gradient is introduced---a spatially sparse, high-frequency supervision signal that provides insufficient gradient diversity to escape saddle-point solutions in the absence of a well-initialised latent space. The result generalises observations from masked autoencoders~\cite{ref15}: a dense auxiliary objective stabilises the latent space before sparse supervision is applied. This failure mode is consistent with theoretical analyses of rank collapse~\cite{ref22} and attention entropy collapse~\cite{ref23} in transformers: both phenomena are exacerbated when input tokens carry little or no signal, as occurs under extreme RSSI sparsity. Standard transformer self-attention has similarly been shown to degrade on spatially sparse data without architectural or training adaptations~\cite{ref24}. Notably, CNN U-Net tolerates single-stage training ($F_1=0.879$), demonstrating that the protocol's necessity is architecture-dependent: convolutional inductive biases implicitly enforce spatial coherence that Transformer encoders must acquire explicitly through the heatmap pretext.

\textbf{Trans. No-Spatial vs. Transformer 2S.} Removing the $4\times4$ spatial scatter and replacing it with global mean pooling yields a marginal performance gain at $N=400$ ($F_1=0.817$ vs.\ 0.792). This counter-intuitive result suggests that at high observation counts the fixed $4\times4$ grid resolution becomes a bottleneck, and the encoder benefits from attending over all tokens globally. At low $N$, the scatter grid supplies an inductive bias that compensates for extreme sparsity.

\textbf{Path-based sampling gap.} Under path-based sampling ($N=48$, single trajectory), the best model reaches $F_1\approx0.25$. Three structural factors drive this gap: (i) spatial coverage asymmetry---50\%--70\% of the room is unobserved, leaving most $4\times4$ scatter cells empty; (ii) coarse grid resolution (2.5~m/cell) that cannot triangulate short walls from unilateral observations; and (iii) tight Gaussian label variance ($\sigma=0.156$~m) providing no gradient signal for predictions deviating by more than a single spatial bin. These are engineering limitations rather than fundamental bounds. Multi-trajectory fusion---combining observations from two orthogonal walks---is a promising direction that requires no model retraining.

\textbf{RSSI vs. CSI.} Our framework deliberately targets RSSI rather than full channel state information (CSI). While CSI provides richer angular and delay information, it requires MIMO antenna arrays and baseband access unavailable to commodity devices. RSSI is available from any single-antenna receiver via standard 5G NR measurement reports, making opportunistic collection feasible at zero infrastructure cost. This positions our approach as complementary to CSI-based ISAC methods~\cite{ref1}, trading sensing resolution for deployment practicality.

\textbf{Limitations and future directions.} Our evaluation is fully simulation-based using Sionna RT~\cite{ref18}, which models specular reflections and diffraction accurately but omits furniture scattering and human-body blockage. The transfer to real deployments will require domain adaptation (e.g., few-shot fine-tuning on measured reference points). The fixed $64\times64$ output resolution may be insufficient for large ($>15$~m) rooms; a multi-scale decoder or tiling strategy is a natural extension. Finally, extending the model to 3D scenes with multiple floors and angled walls would significantly increase practical scope, but requires new simulation data and richer spatial encoding.

%% file: sections/08_conclusion.tex
\section{Conclusion}
We have presented a systematic study comparing single-stage and two-stage training across seven architectures on the first RSSI-to-wall benchmark dataset. The results expose a clear architectural divide: the heatmap pretext is indispensable for attention-based encoders, yet convolutional models perform well without it. CNN U-Net 1S attains Wall $F_1=0.879$, CNN U-Net 2S reaches 0.854, and Transformer 2S delivers 0.792 at $N=400$. Without the pretext, the Transformer encoder degenerates to a constant prediction ($F_1=0.166$), whereas CNN variants remain unaffected. Under path-based sampling ($N=48$), the best model averages $F_1\approx0.24$ across topologies, highlighting the coverage asymmetry of single-trajectory acquisition as the next engineering bottleneck. Future work will target finer spatial scatter grids, multi-trajectory fusion for path-based deployments, sim-to-real domain adaptation, and end-to-end validation on live 5G testbeds.

%% file: references/references.tex
% Bibliography transcribed from the PDF, preserving its numbering and wording.
% No external bibliographic corrections or additional metadata have been added.
% A BibTeX .bib file is not required: these entries compile directly.
% Match the original last-page column break before reference [6].
\IEEEtriggeratref{6}